# SPECTRA: Synthetic IR Test Collections with Relevance Oracles and Controlled Distractor Diagnostics

**Eric Liang**
Oracle
zixuan.liang@oracle.com

**Abstract.** Scalable information retrieval testing needs corpora that are large enough to stress index construction, ranking latency, query routing, and evaluation tooling, yet human-judged test collections remain expensive and may be unavailable when documents are private or still under design. This paper introduces SPECTRA, a reproducible framework for generating synthetic IR test collections through a separation of latent topical structure, surface text realization, metadata controls, query intent generation, and deterministic relevance oracles. Unlike statistical corpus simulators that primarily fit document-length and term-frequency distributions, SPECTRA emphasizes auditable truth conditions, graded qrels, and controlled retrieval-failure diagnostics. The framework is intended as a diagnostic complement to Cranfield-style and TREC-style evaluation, not as a replacement for human assessment. A single-process Python prototype generated corpora up to 60,000 documents and 9.61 million tokens while preserving controllable long-tail vocabulary growth and producing graded relevance labels for 96 queries. In the local simulation study, generation remained close to linear at roughly 12K to 14K documents per second, estimated Zipf slopes stayed near 0.86 in absolute value, and increasing cross-topic distractor text reduced BM25 nDCG@10 from 1.00 at 2% distractors to 0.43 at 36% distractors. These results show that lightweight synthetic corpora can expose retrieval-system scaling and failure modes before costly collection construction begins.

## 1. Introduction

The central difficulty in information retrieval (IR) testing is that the most valuable evaluation artifacts are also the hardest to obtain. A complete test collection needs a document corpus, representative user information needs, query formulations, relevance judgments, and standard metrics. The Cranfield paradigm established the value of controlled test collections for repeatable retrieval experiments [1], and TREC expanded that paradigm into a durable infrastructure for large-scale IR evaluation [2], [3]. Public benchmarks such as BEIR and MS MARCO have further accelerated retrieval research by making large query and judgment sets broadly available [12], [13].

Those resources remain essential, but they do not cover every engineering need. A team building a domain-specific search engine may need to know how an index behaves at 100 million passages before the corpus exists. A retrieval-augmented generation system may require temporal, metadata, access-control, or long-tail entity patterns that are absent from standard benchmarks. A laboratory may need stress tests that can be released without exposing private text. These use cases require synthetic corpora whose statistical properties and ground truth are controlled rather than merely sampled from public data.

Synthetic retrieval collections are not new. Earlier simulation work modeled macro properties such as document length, vocabulary distribution, corpus growth, and query-log behavior [7]. Puranik recently studied efficient generation of synthetic text corpora for scalable IR testing, emphasizing document-length models, Zipfian term generation, singleton modeling, vocabulary growth, and low-memory on-the-fly corpus construction [8]. Recent work also uses large language models to generate synthetic queries, relevance labels, or entire test collections [9]-[11], and query-generation methods such as InPars and Promptagator show that synthetic training pairs can materially improve neural retrieval in low-supervision domains [16], [17]. The gap addressed here is narrower and operational: how to generate a large synthetic IR test collection quickly while retaining an inspectable relevance oracle, query intents, metadata controls, and tunable stress factors that can be used for scalable retrieval diagnostics.

This paper contributes: (i) a requirements model for synthetic IR test collections; (ii) SPECTRA, a simulation architecture that separates statistical corpus controls from latent facts, metadata, surface text, queries, and graded relevance; (iii) a reproducible prototype and local simulation study measuring scaling, corpus statistics, and retrieval degradation under controlled distractors; and (iv) practical guidance for combining synthetic corpora with human-judged benchmarks.

## 2. Related Work

IR evaluation rests on the interaction between test collections, retrieval models, and metrics. BM25 remains a strong lexical baseline because it combines term-frequency saturation, inverse document frequency, and document-length normalization within the probabilistic relevance framework [4]. Graded measures such as discounted cumulative gain and nDCG support evaluation when documents vary in usefulness rather than being simply relevant or not relevant [5]. Reproducible toolkits such as Anserini and Pyserini make these baselines easier to audit across collections and indexing settings [14], [15].

Generative topic models offer one family of corpus simulators. Latent Dirichlet Allocation treats documents as mixtures of topics and topics as distributions over words [6]. Although modern retrieval systems often use transformer representations rather than topic models, the separation between latent semantic structure and observed lexical surface remains useful for simulation. It lets a corpus designer know which documents should be relevant before a ranking model sees only terms, passages, or embeddings.

The closest prior work is Hawking et al.'s simulation program, which explicitly studies emulation of document length distributions, word-frequency distributions, textual representation, corpus growth, query generation, and query logs [7]. Puranik's synthetic-corpus framework makes this statistical simulation line more concrete by comparing document-length models and compact Zipfian term-generation schemes for scalable IR benchmarking [8]. More recent synthetic-test-collection work asks whether LLMs can generate queries, labels, or documents that preserve system-comparison reliability [9]-[11]. SPECTRA is positioned between these lines: it can use statistical corpus controls, but its main object is an end-to-end test collection with query intents, metadata, graded qrels, and deliberately injected distractors. A deterministic simulator can be paired with an LLM realizer, but the test collection's truth conditions do not depend on the LLM remembering its own generated facts consistently.

The framework also draws on adjacent data-management concerns. Metadata harmonization work emphasizes that corpora become more discoverable when structured resource descriptions use consistent schemas [19]. High-cardinality categorical representation matters when synthetic documents carry source IDs, entity IDs, authors, venues, or product identifiers [20]. Automated date-format detection is relevant when corpora include timestamp fields that must be parsed consistently by downstream analytics [21]. Finally, probabilistic certification work motivates reporting uncertainty and confidence bounds for simulation-derived estimates instead of treating a single synthetic run as definitive [22].

# 3. Problem Statement and Design Goals

Let C_N be a corpus of N documents, Q_M a set of M query records, and R a graded relevance relation mapping query-document pairs to labels such as 0, 1, 2, and 3. A synthetic IR simulator receives a control vector theta describing vocabulary size, topic count, topic overlap, document-length distribution, metadata schema, temporal coverage, distractor rate, and query intent mixture. It outputs C_N, Q_M, R, and diagnostic statistics. The goal is not to imitate a particular real corpus perfectly. The goal is to create a transparent testbed whose stress factors are known and whose failure modes can be interpreted.

| Requirement | SPECTRA design choice | Testing value |
| --- | --- | --- |
| Scalable generation | Streaming document construction with precomputed sampling weights and bounded per-document state. | Supports index sizing, throughput testing, and repeated parameter sweeps. |
| Known relevance | Latent topics and anchor terms define a deterministic graded oracle before ranking. | Allows repeatable nDCG, recall, and ablation studies without human labeling cost. |
| Metadata control | Documents include genre, date, visibility, and source-like categorical fields. | Tests faceting, filters, temporal freshness, and high-cardinality routing. |
| Tunable difficulty | Cross-topic distractor text and topic overlap create lexical ambiguity. | Exposes scoring assumptions and brittle query-time heuristics. |
| Auditability | All parameters, random seeds, qrels, and diagnostics are exportable. | Makes synthetic results reproducible and separable from model claims. |

*Table 1. Design requirements for synthetic IR test corpora and their corresponding implementation choices.*

The most important design constraint is auditability. A synthetic corpus may contain readable text, template text, or purely symbolic tokens, but it should expose the reasons a document is relevant. Otherwise, it risks becoming an opaque benchmark whose biases cannot be diagnosed. For that reason, SPECTRA treats surface realization as a replaceable layer rather than as the source of ground truth.

# 4. The SPECTRA Framework

SPECTRA, short for Synthetic Probabilistic Evaluation Corpus for Text Retrieval Assessment, has five layers. The latent layer samples topics, entities, dates, and document roles. The surface layer turns those latent assignments into text. The query layer samples information needs and query formulations. The oracle layer converts latent relations and observed anchor overlap into graded qrels. The export layer writes documents, queries, qrels, run files, and diagnostics in formats that can be consumed by standard retrieval tooling.

In the prototype used here, surface text consists of synthetic lexical tokens rather than natural sentences. This choice is deliberate: the experiment targets indexing, ranking, and relevance-control behavior, not human readability. A production version could replace the token realizer with templates or LLM-generated prose while preserving the latent oracle. The boundary between oracle and surface realizer is the safeguard that keeps synthetic corpora verifiable.

For each document, the generator samples a dominant topic, a secondary topic, metadata fields, and a document length from a bounded log-normal distribution. Tokens are drawn from four pools: common terms, dominant-topic terms, secondary-topic terms, and long-tail terms. A distractor parameter controls how much vocabulary from an unrelated topic is injected. Queries are produced from topic anchor terms, with optional metadata intent. A document receives a high relevance grade when its latent topic matches the query and it contains multiple query anchors; secondary-topic matches and lexical-only overlaps receive lower grades.

The framework is intentionally compatible with existing retrieval stacks. Documents can be exported as JSONL for Lucene-based systems, TSV for dense retriever pipelines, or compact postings for fast prototype experiments. Relevance labels follow the same conceptual contract as qrels used by TREC tooling. This allows synthetic data to serve as a pre-benchmark diagnostic stage before experiments move to BEIR, MS MARCO, TREC tracks, or private judged collections [2], [3], [12], [13].

# 5. Simulation Study

## 5.1 Experimental Setup

The local prototype was implemented in Python using deterministic pseudo-random seeds. It used 32 latent topics, a 120,000-token vocabulary, 96 queries, a common vocabulary band, topic-specific vocabulary bands, and a long-tail vocabulary. The retrieval baseline built an in-memory inverted index and evaluated BM25 against a raw term-frequency baseline. The study reports nDCG@10 and recall@100 over the generated graded qrels. All measurements were obtained from a single-process prototype in the local workspace; they should be interpreted as diagnostic measurements rather than as optimized systems results.

Two experiments were run. The scaling experiment used 5,000, 20,000, and 60,000 documents with a fixed 12% cross-topic distractor rate. The difficulty experiment used 20,000 documents while varying distractor rate from 2% to 36%. The simulator also estimated corpus diagnostics: average document length, vocabulary size, postings count, a log-log Zipf slope over the most frequent terms, and a Heaps-style vocabulary growth exponent over generation checkpoints.

| Docs | Tokens (M) | Gen kdocs/s | Tok M/s | Vocab | \|Zipf\| | Heaps | BM25 nDCG@10 |
|---|---|---|---|---|---|---|---|
| 5,000 | 0.79 | 12.36 | 1.95 | 47,388 | 0.86 | 0.43 | 0.986 |
| 20,000 | 3.19 | 12.65 | 2.02 | 85,806 | 0.86 | 0.43 | 0.994 |
| 60,000 | 9.61 | 13.42 | 2.15 | 111,912 | 0.86 | 0.34 | 0.996 |

*Table 2. Scaling results for the synthetic corpus simulator at a fixed 12% distractor rate.*

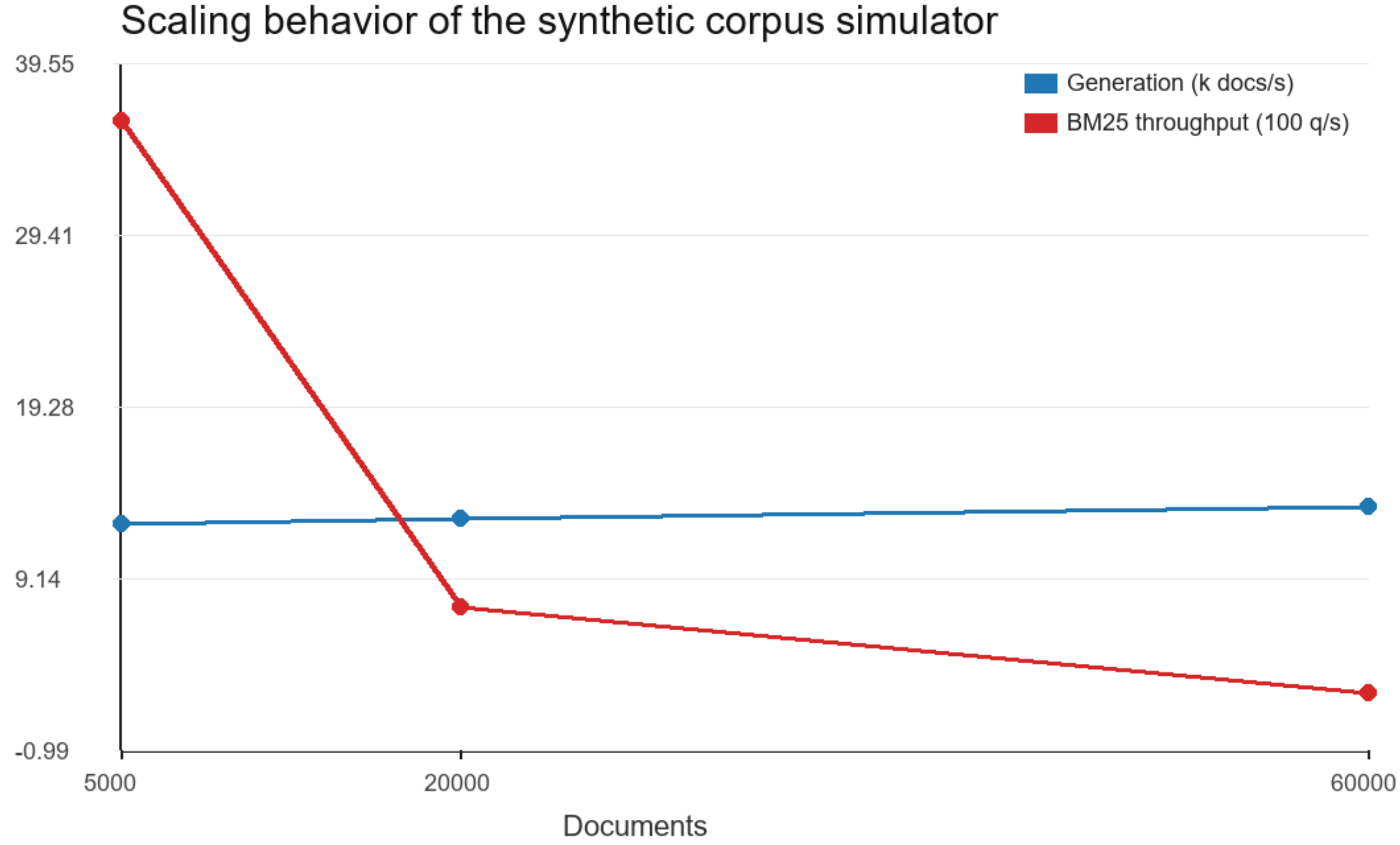

*Figure 1. Generation throughput remains near-linear while query throughput falls as postings grow. BM25 throughput is shown in hundreds of queries per second for scale comparability.*

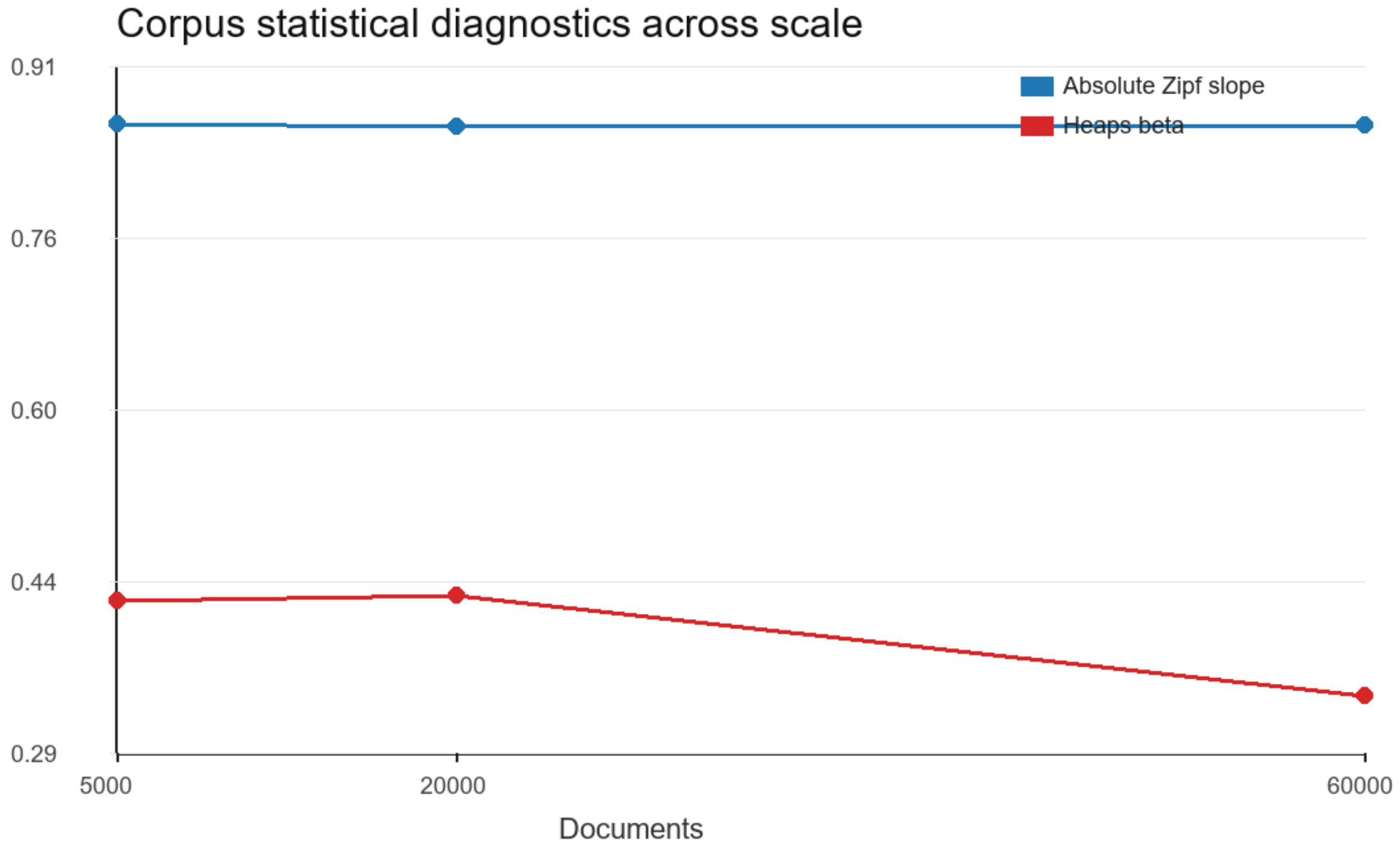

*Figure 2. Corpus statistics remain stable across scale: absolute Zipf slope stays near 0.86, while Heaps growth decreases at the largest size as the finite vocabulary becomes more covered.*

## 5.2 Difficulty and Retrieval Behavior

| Distractor | BM25 nDCG@10 | TF nDCG@10 | BM25 R@100 | TF R@100 | Median rel. docs |
|---|---|---|---|---|---|
| 2% | 0.999 | 0.968 | 0.159 | 0.158 | 666 |
| 12% | 0.990 | 0.960 | 0.155 | 0.159 | 674 |
| 24% | 0.838 | 0.898 | 0.127 | 0.140 | 666 |
| 36% | 0.434 | 0.216 | 0.083 | 0.069 | 656 |

*Table 3. Retrieval quality under increasing injected cross-topic distractor text.*

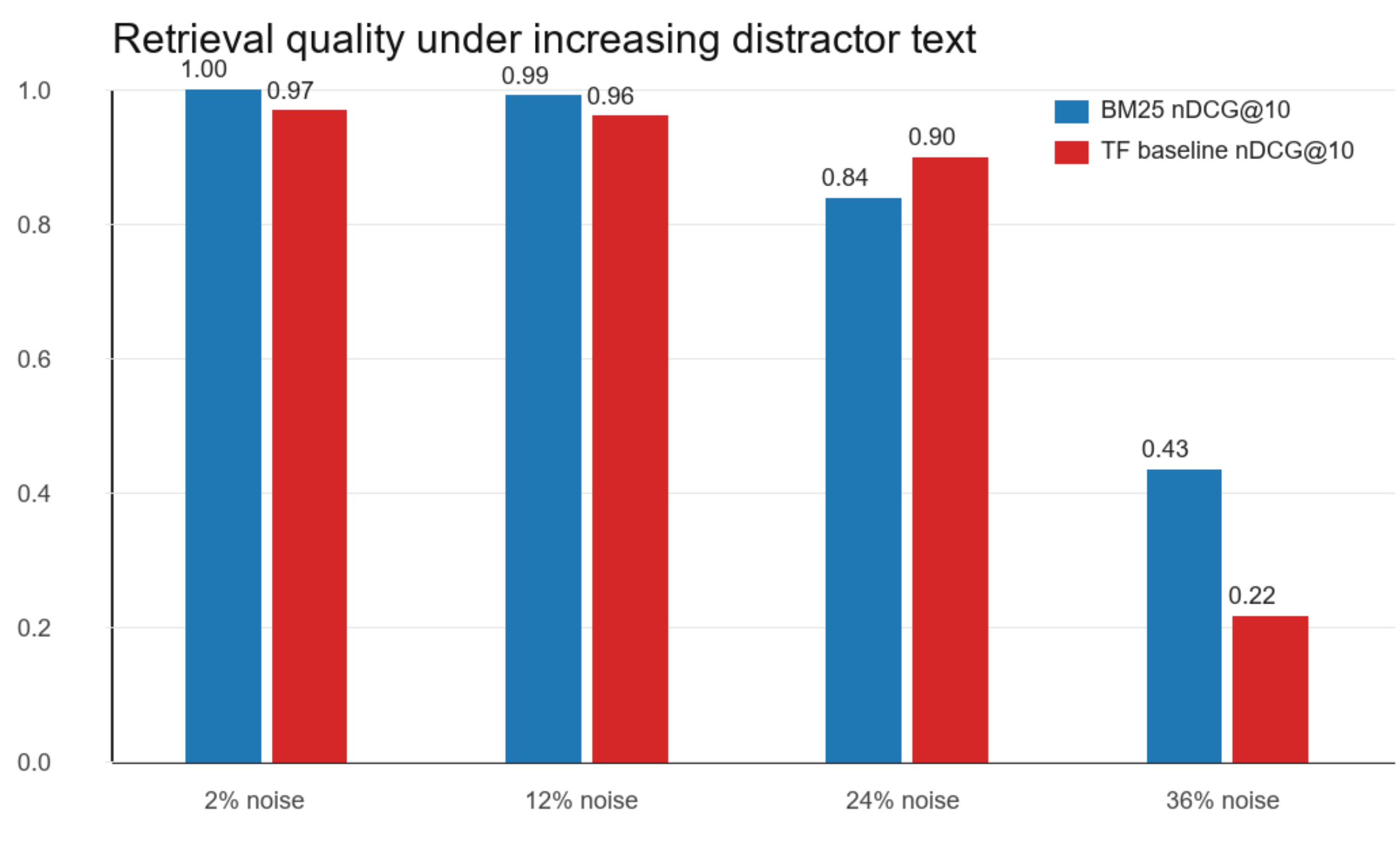

*Figure 3. Controlled distractors make the synthetic test collection progressively harder and expose differences between BM25 and a raw term-frequency baseline.*

The difficulty experiment behaves as intended. BM25 is nearly perfect when distractor vocabulary is rare, remains high at 12%, and then degrades sharply as unrelated documents contain more query-topic anchors. The raw term-frequency baseline overtakes BM25 at the 24% distractor condition, which is not a contradiction. In that setting, repeated anchor terms become an artificially strong signal, while BM25's saturation and length normalization reduce the reward for repetition. A synthetic testbed is useful precisely because such scoring assumptions can be isolated.

Recall@100 decreases with scale because the number of relevant documents per query grows while the cutoff remains fixed. This is a useful reminder for scalable testing: a metric can change because the collection changed, not because the ranking system changed. Synthetic corpora make that effect observable by exposing the number and grade distribution of relevant documents for every query.

## 6. Discussion

The results support three claims. First, a deterministic simulator can generate millions of tokens quickly enough for repeated test sweeps. Second, latent-oracle design makes relevance labels cheap without hiding why they exist. Third, controlled distractors provide a direct way to stress ranking assumptions. These claims are deliberately modest. The prototype does not show that symbolic synthetic text has the same evaluation validity as human-written corpora. It shows that such text can be useful for engineering diagnostics before more expensive evaluation begins.

The most promising use case is staged evaluation. A search team can first use SPECTRA to verify that indexing, sharding, metadata filtering, freshness ranking, qrel handling, and metric computation work at target scale. Then the team can move to public or private judged collections for validity. This staging mirrors good software testing practice: synthetic corpora are unit and stress tests for retrieval infrastructure, while human-judged benchmarks remain the integration test for user value.

LLM-generated text can be incorporated, but the simulator should keep a ground-truth event bus or latent record separate from the generated prose. This separation reduces the risk that generated documents contradict each other, leak private exemplars, or create untraceable labels. When LLMs are used, the latent layer can provide structured prompts, entity constraints, and negative-control documents; the oracle can then validate that generated passages still express the intended facts.

Metadata is not decorative in scalable retrieval testing. Real search systems often route, filter, rank, and monitor by source, date, user group, document type, language, or authority. Synthetic corpora should therefore include high-cardinality and temporal fields, not only bag-of-words text. The metadata harmonization, categorical representation, and date-detection work cited above provide useful building blocks for making these fields analyzable rather than incidental [19]-[21].

## 7. Limitations and Future Work

The main limitation is surface realism. The local experiment used synthetic tokens, so it cannot test passage semantics, entity grounding, discourse coherence, multilingual morphology, or dense embedding behavior in a realistic way. Future work should add template and LLM realization modes, then compare system rankings against public benchmarks to measure when synthetic rankings transfer. A second limitation is that the oracle is only as good as its design. If the simulator encodes a narrow notion of relevance, systems can overfit to that notion. Future versions should support multiple relevance theories, assessor disagreement models, and confidence intervals over synthetic runs.

The prototype also does not yet export full Lucene/Pyserini-compatible collections, dense embeddings, or query logs with sessions. These are engineering extensions rather than conceptual blockers. The most important next step is calibration: fit simulator parameters to target domains while preserving privacy. Calibration should report what is matched, such as length distribution, vocabulary growth, metadata cardinality, query length, and topical overlap, and what is intentionally abstracted away.

## 8. Conclusion

Synthetic text corpora can make scalable IR testing cheaper, faster, and more reproducible when they are treated as diagnostic instruments rather than substitutes for human judgment. SPECTRA provides a concrete architecture for doing this: generate latent topics and metadata, realize surface text, produce queries, derive graded relevance from an auditable oracle, and export diagnostics alongside the collection. The local prototype demonstrates that this approach can generate and evaluate large controlled corpora quickly while exposing retrieval failure modes through tunable distractors. The broader opportunity is to combine synthetic scale with human validity, using each where it is strongest.

## Data and Code Availability

The deterministic prototype used for the simulation tables and figures is included in the project workspace as run_simulation.py together with the generated JSON results and chart images. The prototype does not use external corpora or private data.